# Microscopic Observation of the Light-Cone-Like Thermal Correlations in Cracking Excitations

H.O. Ghaffari [1(a)], W.A. Griffith

[1] *University of Texas at Arlington, Box 19049, Arlington, TX, 76019, USA*

Many seemingly intractable systems can be reduced to a system of interacting *spins*. Here, we introduce a system of *artificial acoustic spins* – fictitious spins which are manipulated with ultrasound excitations associated with micro-cracking sources in thin sheets of crystals. Our spin-like system shows a peculiar relaxation mechanism after inducing an impulsive stress-ramp akin to splitting, or rupturing, of the system. Using real-time construction of correlations between spins states, we observe a clear emergence of the light cone effect. It has been proposed that equilibration horizon occurs on a local scale in systems where correlations between distant sites are established at a finite speed. The observed equilibration horizon in our observations defines a region where elements of the material are in elastic communication through excited elementary excitations. We demonstrate that prior to the applied stress-ramp, the system is described by an algebraic correlation function, and above the critical point the correlation function shows exponential scaling characterized by the formation of kink-pairs. These results yield important insights into dynamic communication between failing elements in brittle materials during processes such as brittle fragmentation and dynamic stress triggering of earthquake-generating faults.

**Introduction-** Cracking or splitting of a many body system is accompanied by excitations, interactions and recombination of many quasiparticles, and thus represents an extremely challenging problem [1-5]. The protocol of splitting, commonly referred to as rupture, in brittle solids is believed to be controlled in a process zone very near to the moving rupture tip. On the other hand, sudden injection of many emitted quasi-particles strongly induces a fast change in the state of the system yielding out-of-equilibrium evolution which eventually relaxes to a steady state [6-8]. To establish a steady state regime, the elements of the system must communicate with each other: the distant elements of the system must be correlated. Based on this requirement, then, the propagation of real-time correlations is a topic of utmost relevance in the study of fractures and fragmentation (multiple splitting) [9-10].

The spread of correlations in space-time is due to propagation of quasi-particles pairs forming light-cone-like propagation which establish the equilibration horizon [11-14]. In this work, we use a novel technique to study the emission of phonon excitations in relatively strong mechanical cracking sources (i.e., acoustic crackling noises) during indentation of thin films of natural occurring minerals. We use multi-array high frequency acoustic phonon sensors arranged in a ring and map their recorded excitations to a classical spin-like system with nanosecond resolution. We show that cracking-induced ultrasound excitations manifest themselves in flipping of the fictitious spins. We demonstrate that the system crosses a transition point below which the correlation function shows power law scaling decay, and above which the initial exponential function holds. Furthermore, our results establish that above the transition point, kink-pairs are formed, and these topological defects are associated with the exponential form of the correlation function. These results affirm that the relaxation of the system has already begun during the kink-pair generation. We also recognize slow fronts under slow stress-ramps which propagate at an order of magnitude slower than the fast fronts. We pay particular attention to the transitions from slow to fast fronts as well as the bifurcation and proliferation. Our results open horizons for understanding incipient plasticity in microscopic dynamic failure, and, more generally, non-equilibrium relaxation of many body systems.

**Results-** Our experiments include micro-indentation of thin sheets of two different minerals (muscovite mica and graphite) at room temperature, and the corresponding material failure is recorded as multi-array ultrasound excitations (see Methods section for details of the sample preparations and data acquisition). Different loading protocols are employed to indent the thin films, and all data are recorded with resolution of 40MHz. In Fig.1, we show multi-array waves from a single cracking excitation parallel to the [001] face of a *muscovite mica* sheet. During a central-indentation test of the mica sheet, we were able to record up to 10 excitations with different excitation levels (i.e., cracking energy). By mapping this system to a classical spin system, we elucidate a new phenomenon in this relatively simple experiment: a clear spread of a light-cone-like thermal correlation front. To map our ring-like structure with finite nodes to spin-like particles, we use a thresholded-measure of each node's activity (measured in *mV*) to construct a network of nodes per each recorded data point (see Methods section for details of the algorithm). The constructed graph is characterized by the average of all nodes' degree (<k> -as well as many other measures [15-18]) and evolution of these graphs help to unravel the complexity of the excitation sources [16-18]. The degree $k_i$ of the $i^{th}$ node at a given time represents the number of connected links to the node where "links" are established based on a similarity metric. Later, we will see that <k> may be viewed as playing the role of a tuning parameter to induce transition. To visualize the evolution of the system, we map the spatial evolution of the degree $k_i$ of the $i^{th}$ node (Fig.1d), using polar coordinates $(r_i, \theta_i)_{i=1,...,Nodes}$ where $r_i = k_i$ and $\theta_i$ indicates the position of the node around the ring (Fig.1d-we call this structure K-chain). Next, for each node we assign $s_i = sign(\frac{\partial k_i}{\partial t})$ and then $s_i = \pm 1$. With this mapping, each node in a given time step acquires one of the states (↑ or ↓). This can be interpreted as the surface of a constant phase in analogy with the wavefronts and defects of this surface (i.e., defects in network structure) bear a resemblance to *Nye-Berry's theory* on dislocations in wave trains [31]. Here we use a linear chain simplification but the true nature of chains is 3d where the state of each node can be represented with a 3d vector such as $\vec{S} \propto \sin\Theta[\cos(\Phi)\hat{x} + \sin(\Phi)\hat{y}] + (\cos\Theta)\hat{z}$ with $\Theta$ as the polar and $\Phi$ as azimuthal angles. In a linear *x*-chain configuration, we restrict ourselves to $\Theta = \pi/2$, $\Phi = \pm\pi/2$ and $|\vec{S}| = 1$.

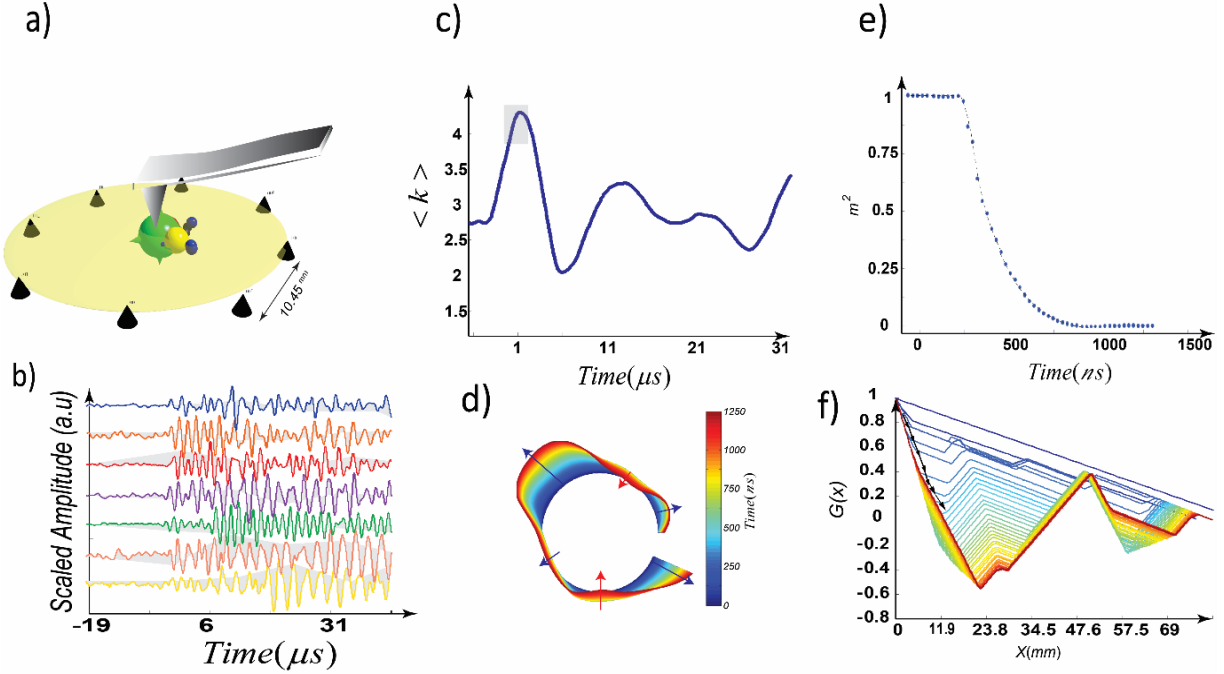

**Figure 1| indentation experiments and sample preparation a)** Schematic representation of indentation test acting at the center of a suspended ultra-thin mica sheet. Ultra-thin Mica sheet is suspended over 8 pico-sensors with diameter of ~4.77 mm and height of ~3.9 mm. The spheres correspond with the locations of recorded emitted acoustic phonons. **(b)** Scaled recorded multi-array acoustic waveforms in tens of μs for the highlighted event as the green-sphere. **c)** Evolution of average of all nodes' degree <k>, **d)** Accumulated K-chain patterns in ~1250*ns* corresponding to the time range indicated by the gray box in *c)*, **e)** The *magnetization squared*-like parameter versus time extracted from *K-chain's* evolution. The system effectively is *frozen* around *t=1000 ns* for ~150*ns* implying a non-adiabatic regime due to fast (local) stress ramp. **f)** The real-time correlation functions for the shown example. The correlation fronts, on average, spread with well-defined velocity of 5.05km/s (the arrows show the cross-over distances).

Our system behaves with a universal pattern characterized by the following steps: (1) the system is pulled to a state where all spins polarize in the outward direction. This is the first, rising part of <k(t)> in Fig.1c and is characterized by $m=<s>=1$ (Fig.2c), and we refer to this stage as the "ordered phase". Our analysis shows that this phase is stable for ~200-500ns. We fit a two-point correlation function in the form of $G(x) \equiv (1-\frac{x}{L})\exp(-\frac{x}{\xi})$, where $L$ is the total number of spins, $x$ is distance, and $\xi$ is the correlation length as the cut-off length of the correlation function where for distances shorter than the correlation length, G(x) can be fit by a power law function (Fig. 1f). For the aforementioned ordered phase (the first step of the universal pattern), a triangular function is given where $\xi \to \infty$ (i.e., algebraic scaling- Fig.1f and Fig.2a-b). (2) The system is suddenly stress-ramped (i.e., <k> ramp) and during the ramp the spins are flipped. We can see the signature of this fast ramp in a space-time correlation map (Fig.2a). The fast ramp results in flipping the sign of individual spins which results in "folding" of the chain (Fig.1d, Fig.S3 and Fig.S5). Careful monitoring of the evolution of the correlation function indicates that the onset of the fast ramp yields a rapid break-down in the initial triangular function. We refer to distance at which this break down occurs at each time step as the cross-over distance (Fig. 2d). At distances smaller than the cross-over point the system *forgets* the initial long-range coherence, and beyond this distance the correlation approximately recovers its initial triangle trend (Fig.1f-Fig.2d and Fig.3).

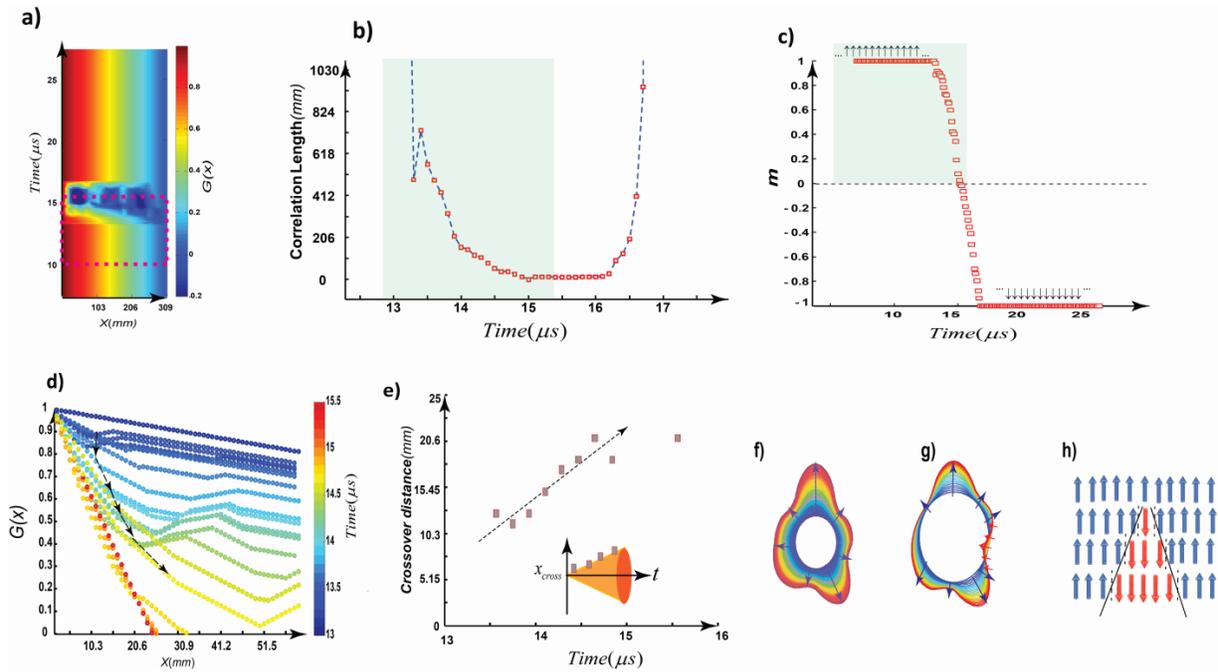

**Figure 2| Real-time dynamics of Correlations as a function of time and space. (a)** The stress-ramp rapidly changes the initial state of the system. We show the space-time two-point correlation function in Graphite. The time interval of interest is indicated by the dotted rectangle. **(b,c)** The correlation length decays from an infinite value to an effective stable regime characterized by the order parameter, *m*, approaching zero. **(d,e)** The correlation functions in each time step are used to follow the spread of the correlation front (the arrows show the cross-over distances). The decay of correlations is characterized by a front moving with a finite velocity. In **e**) we show the positions of the cross-over distance as a function of evolution time *t*, revealing the *light-cone-like* decay of correlations. The solid line is a linear fit, the slope of which corresponds to twice the characteristic velocity of front correlations. **(f-h)** fully polarized spins are locally perturbed with flipping a site which induces propagation of fronts in left and right hands from the disturbed node.

The result of this procedure is shown in Fig. 1f for Mica and Fig.2d for Graphite. We observe a clear, almost linear, scaling of the cross-over position which is characterized by a slope corresponding to twice the velocity of correlations (Fig. 2e). Examining this characteristic velocity for at least 10 tests with similar ramp rates indicates that average propagation velocity of the correlation front is ~ 4.32±.5 km/s for Graphite film and 5.05±0.6 km/s for muscovite mica. The velocities of fronts are approximately the same as, or below, the longitudinal sound velocity of the films (Methods section and Fig.S7) proving that when the system is ramped, acoustic phonons are generated rapidly. The successive cross-over distances can be approximated by a simple linear function (see Fig.1f –Fig.2e). Ideally, the perturbation of the fully polarized state for a single flipping - prior to multiple active excitations - can be understood as the creation of a phonon and then with the associated phonon dispersion relation [19-22].

Further intriguing observation is the emergence of kink-pairs which we clearly see in terms of folding of visualized links all over the ring (Fig.1d), a process that occurs during the stress-ramp. The emergence of kink-pairs induces a more disordered phase field, which itself yields exponential scaling of the correlation function (i.e., decreasing correlation length-Fig.2b). The correlation function shows approximately an oscillatory behavior indicating that kinks are anti-correlated and their final distance keep more or less the same distance (i.e., frozen characteristic length) from each other (Fig.3) [23-24]. The generation of kink-pairs can thus be characterized by the magnetization squared-like parameter $m^2 = (\frac{1}{N}\sum_{i=1}^{N} S_i)^2$ where $N$ is the number of spins. In Fig.1e, we show the effect of the stress-ramp on decay of $m^2$ where around the peak of $<k>$, $m^2$ is effectively "*frozen*". The same phenomenon occurs for correlation length as depicted in Fig.2b. This frozen regime can be explained with the Kibble-Zurek mechanism where the number of defects is scaled with the stress-ramp duration Fig.2b & c [9, 14, 16, and 24].

The evolution path of the system after crossing the peak of <k> is reversed and the system approaches another degenerate state in which all nodes point inward. Evaluating the path of the system in the emitted signal energy versus <k(t)> space represents a fast change in the signal power resembling (partial) avoided crossing in a driven two-level system [25-27] which is governed by the ramp parameter (Fig.4; Fig.S.1-2). Here we see that the fast ramping causes sweeping from the low energy level to the high energy level. This occurs by generation of topological defects which, based on diabetic two-level system transition, and occurs non-adiabatically around the avoided interval [24-27]. In other words, in the transitions between the states ↑↑...↑ to ↓↓...↓, an ultrasound field with variable frequency is generated which is manifested in a triangle-like pulse (Fig.4a-approximately the shape of <k(t)> as the tuning parameter).

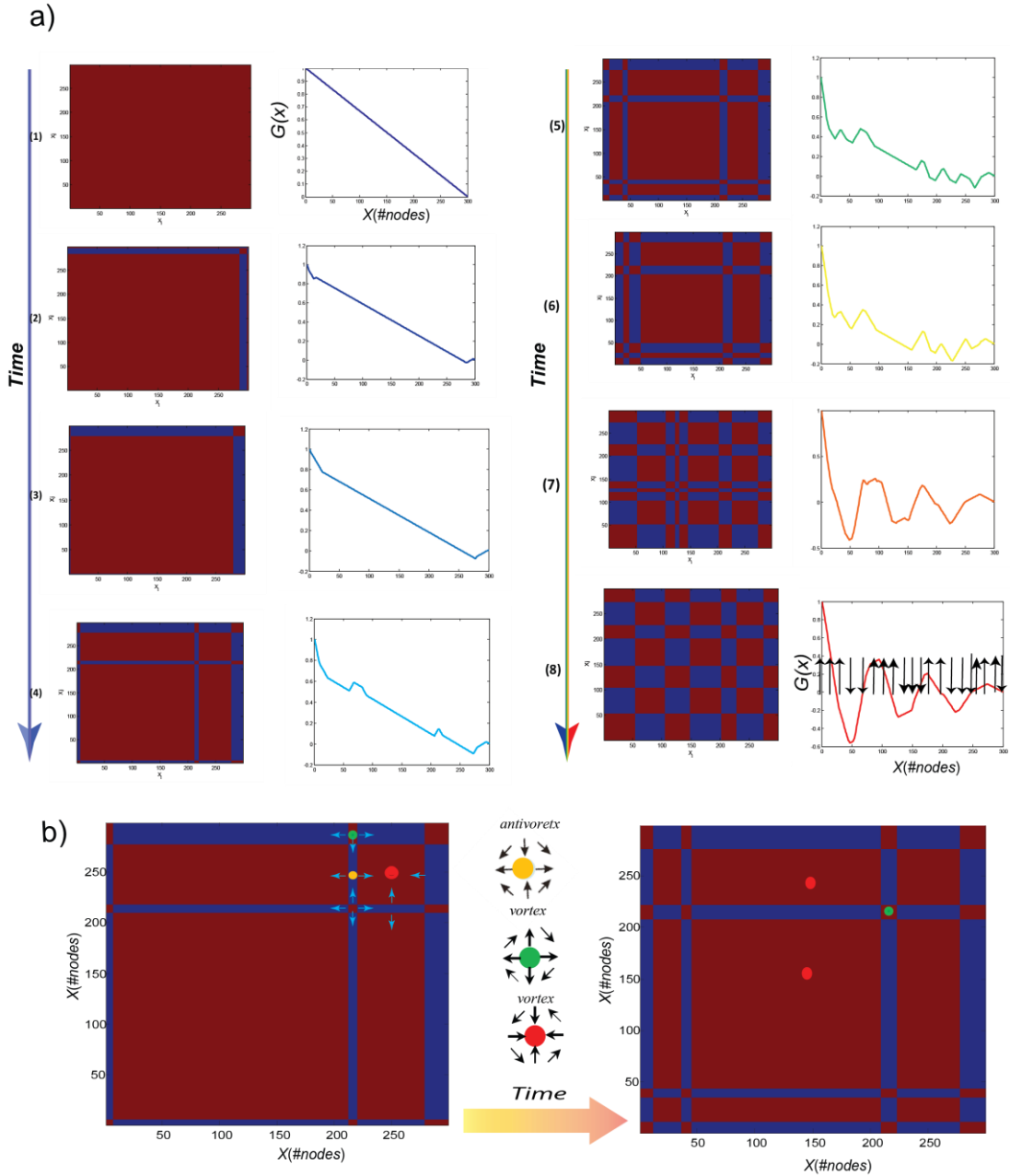

**Figure 3 | Domain wall dynamics.** Snapshots of real-time dynamics of two-point correlation matrix evolution with time (1 to 8) for a single acoustic emission event in indentation of a graphite film. In evolution from snapshots (7) to (8), we observe domain wall dynamics resulting combination and annihilation event. The blue domain is "melted" with combination of two red domains. (b) Shows the source mechanisms of domains. The arrows represent the in-plane components.

This finding in our cracking experiments implies a new understanding of microscopic dynamic failures: let us assume that the ramp profiles (<k(t)>) are reminiscent of the stresses [16-17] which build up to the maximum tolerance point of the system. When the system cannot follow the induced ramp adiabatically, it will cross an energy barrier regime yielding topological excitations. Otherwise the system will slowly follow its adiabatic path to lower or upper energy branches.

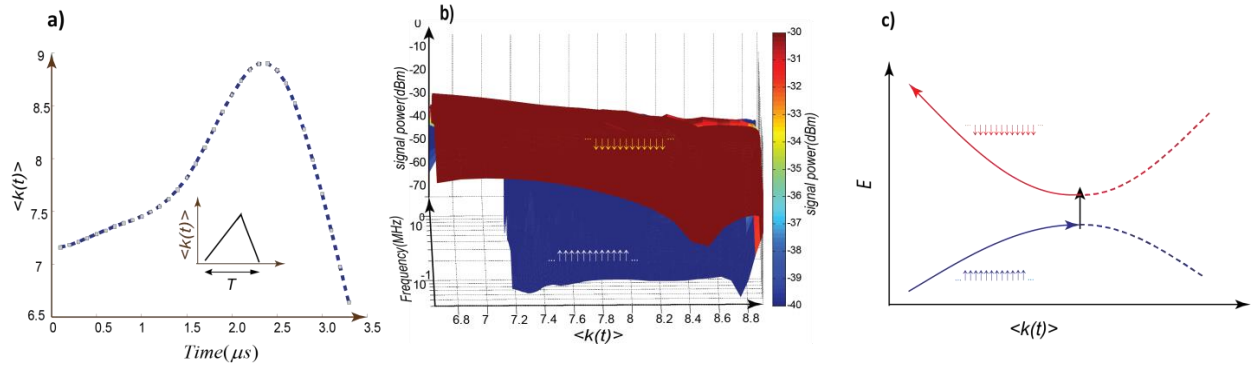

**Figure 4| a,b)** Fast crossing from the low energy to high energy level is governed by tuning parameter <k> . In **(a)**, we have shown the tuning parameter which can be approximated by a triangle pulse (inset). **(b,c)** The transition from ↑↑...↑ to ↓↓...↓ is tuned by <k(t)> and crossing to another level is accompanied by the precipitation out of the topological defects . In (b), we have shown the 3D representation of ultrasound power-frequency-<k(t)> . In **(c)** we show the schematic of the ideal two -level system where in our case studies <u>do not flow</u> in the dotted-lines regime (Also see Fig.S2).

Next, we recognize slow propagation of local correlation fronts up to one order slower than the sound velocity (Fig.5). In this case, the slower reverse and rising ramps, as shown with the blue lines in Fig.5a and c, respectively, are slower than the main rising ramp (red line) by 3 orders of magnitude ($\sim 10^3 s^{-1}$ versus $\sim 10^6 s^{-1}$). The longer ramp time is coincident with the growth of additional maximum, or enhanced, correlations (Fig.5d). These fronts might proliferate in multiple schemas, bifurcate, annihilate or recombine, in parallel with the physics of topological defects (Fig.3 and Fig.5e) [28-31]. The slowest persistent front we could record was spreading at ~625m/s (Fig.5b-front II) following the relatively fast front (~3000m/s – front I). It is noteworthy that the analogy of this observation with the similar propagation of "slow rupture" fronts is reminiscent of slow laboratory earthquakes on frictional interfaces [32-35]. In the case of these laboratory earthquakes, the initial rupture along a frictional interface moves an order of magnitude slower than the Rayleigh wave speed, followed by a fast rupture.

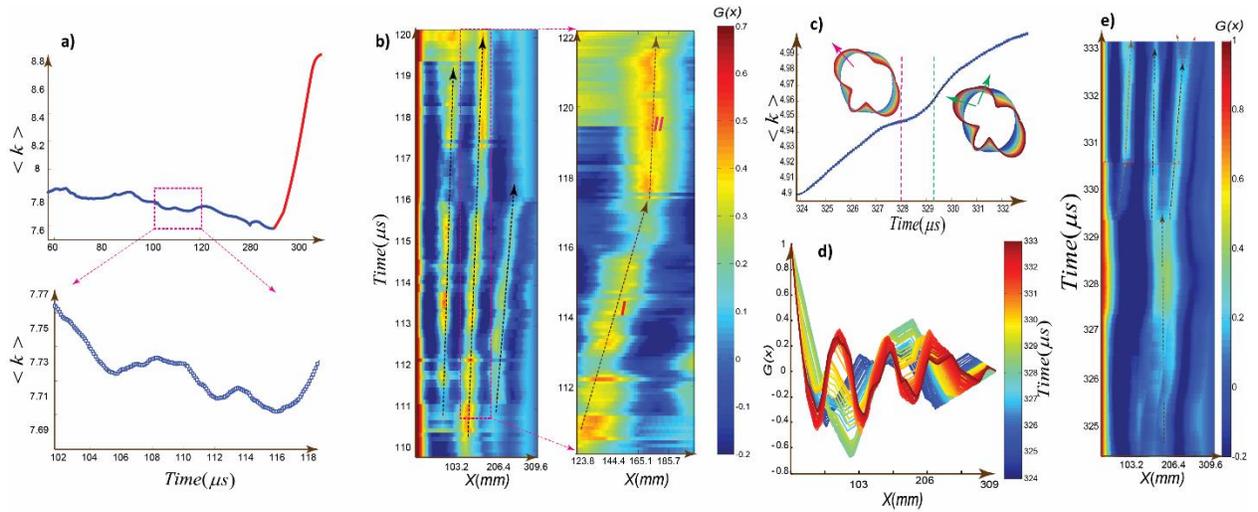

**Figure 5| Propagation of slow correlations fronts. a)** A reverse ramp (right-hand of the triangle pulse in inset of Fig.3a) with much slower rate results the growth of correlation fronts with slow prorogation rate. In this example, we observe proliferation of three fronts with transient velocity from ~1500m/s to ~625m/s in transition from front I to II (**b**). **c)** Another slow ramp with bifurcating fronts as shown in (**e**) at ~329.3μs and ~332.5 μs. A gliding-like dislocation is evident at ~330.5 μs. In this case, the bifurcation event is induced by sudden acceleration in the ramp. The evolution of the K-chain with color assisted code as local time clearly shows the splitting the front (pink arrow). In (**d**) we have shown the correlation function in each time step manifested in space-time correlation in (**e**).

**Discussion-** In summary, we studied the dynamics of fictitious spins constructed from ultrasound waves generated by cracking excitations due to indentations of thin sheets of materials. The pseudo-spin systems were globally swept from a degenerate state to another degenerate state with higher energy. By monitoring the real-time evolution of correlation functions, we found that the dynamics of the correlation functions show a clear "*light cone*"-like behavior: it takes a finite time for the correlations to travel, proportional to the distance between points. The finite travel velocity of the equilibration horizon moves with a well-defined velocity, which are different for the employed thin-films and are at or below the range of the longitudinal phonon velocity. The observed equilibration horizon defines a region (in a given time step) where elements of the material are in *elastic* communication through elementary excitations. We showed that the transition from algebraic to exponential scaling is the result of the generation of kink pairs and the subsequent the relaxation path to a (quasi) stable state in our system is accompanied by removal of kinks. Furthermore, under the slower stress ramp (i.e., <k> ramp) we observed more complex patterns of the correlation fronts including transitions in front velocity, slow propagation and bifurcations. The observed branching of the correlation fronts coincides with the fast switch from the slow to rapid stress ramp. We compare this observation with a macro-scale picture of the bifurcation of ruptures of frictional interfaces which showed that acceleration of slow rupture fronts bifurcate the moving tip into two moving fronts [32-34]. We might understand the simultaneous emergence of slower correlation fronts with fast fronts in terms of multiple stages of relaxations with different timescales. Such relaxation mechanisms have been predicted to happen in many other systems [36-37].

Furthermore, given the resemblance to dynamic evolution of frictional interfaces, we propose that an initial fast relaxation mechanism(s) with light-cone like evolution controls the fast weakening regime in terms of frictional evolution which in turn corresponds to a dynamic stress drop. In this context the results of this study may be applied to understanding weakening mechanisms associated with ruptures in so-called laboratory earthquakes [32-35]. While we showed that our acoustic spin system imprints a light-cone like propagation of information through the elastic waves, we left with the question of the role of interaction ranges between sites on the front velocities. It has been shown numerically and experimentally [11-14] that finite range interactions are associated with a maximum velocity of information propagation and information cannot propagate at an arbitrary speed (Lieb and Robinson's theory [38]). However, it is possible that the system exhibits long-range interactions (i.e., van der Waals interactions in Graphite and Mica). It has been predicted that with increasing interaction range, the maximum velocity of information propagation in the system will diverge, and the light-cone picture is violated [39-40]. It would be interesting to study the effect of long-range interactions on the speeds of the correlation fronts in excitations including 2D structures such as Graphene.

Another interesting study could focus on the role of other quasi-particles and their interaction with emitted phonons. While we focused on the role of non-interacting phonons, it is known that splitting a brittle solid can lead to the emission of high energy electrons and photons [3–4]. This is a crucial point in modern studies of fractures which involve the detaching of chemical bonds, a process for which the true physics is captured by quantum mechanics methods. In addition, emissions of different quasi-particles as well as their interactions complicate the precise elastic communications. Finally, the introduced artificial acoustic spins could be extended to studies of higher-order correlation functions and the 3d version of K-chains with access to phase profiles.

## Materials and Methods

**Experimental Procedures:**

We indented three different crystalline materials (Muscovite Mica, Graphite, Gypsum ($CaSO_4 \cdot 2H_2O$)) and an amorphous glass film. Our focus on the reported tests in the main text was on mica and graphite films (from Princeton Scientific Corp). The indentation on mica was a (001)-oriented mica sheet. We first suspended a thin sheet of mica on a ring–like structure of ultrasound sensors and carefully glued on sensors. Then, we mechanically exfoliated by peeling sheets from the top of the specimen similar to making graphene from graphite [41]. We repeated this process 10 to 15 times to achieve a thin layer of the mineral sheet suspended on the sensors, and we repeated this procedure on large samples of natural Gypsum. The central indentation was performed using a micro-indentation instrument (LM300AT). The minimum and maximum employed indentation loads were 90mN and 490mN, respectively. In order to detect ultrasound excitations we utilized piezoelectric sensors. In the primairy set-up (Fig.1), we used 8 nodes (sensors), while to confirm the results we also employed a second set-up with 16 sensors. The acoustic excitation signals are first pre-amplified at 60 dB, before being received and digitized. The p-wave velocities of specimens are measured prior to indentation tests (and prior to the exfoliation of the samples when the initial thickness was ~500 μm) with pre-defined a 100 V high-frequency (0.5 μs rise-time) pulse which excites each channel while the other channels are recording. Knowing the onset time of the excitation pulse and the sensors positions, the measurement of the travel times yields the average velocity along each ray path.

**Artificial Acoustic spins with ultrasound waves of crackling noises:**

To construct an artificial acoustic spin, we map the recorded multi-array acoustic emission (multiple time series for each event) to a mathematical graph in a way that linked sites form a surface with the same sign (+/-) of the evolution rate. This can be interpreted as the surface of constant phase in analogy with "wavefronts" and defects of this surface (i.e., network structure) sets in parallel with *Nye-Berry's theory* on dislocations in waves [31].

We use a well-established algorithm to construct the mathematical graphs from our reordered acoustic emissions with a fixed number of nodes [16-17, 35]. The main steps of the algorithm are as follows:

(1) The waveforms recorded at each acoustic sensor are normalized to the maximum value of the amplitude at that node.

(2) Each time series is divided according to maximum segmentation, such that each segment includes only one data point. The amplitude of the *j*th segment from the *i*th time series ($1 \leq i \leq N$) is denoted by $u^{i,j}(t)$ (in units of mV). $N$ is the number of nodes or acoustic sensors. We set the length of each segment as a unit with a resolution of 25ns.

(3) $u^{i,j}(t)$ is compared with $u^{k,j}(t)$ to create links between the nodes. If $d(u^{i,j}(t), u^{k,j}(t)) \leq \zeta$ (where $\zeta$ is the threshold level discussed in the following point) we set $a_{ik}(j)=1$ otherwise $a_{ik}(j)=0$ where $a_{ik}(j)$ is the component of the connectivity matrix and $d(\bullet) = \|u^{i,j}(t) - u^{k,j}(t)\|$ is the employed *similarity metric*. With this metric, we simply compare the amplitude of sensors in the given time-step. The employed norm in our algorithm is the absolute norm.

(4) Threshold level ($\zeta$): To select a threshold level, we use a method introduced in [35 and references therein] that uses an adaptive threshold criterion to select $\zeta$. The result of this algorithm is an adjacency matrix with components given by $a(x_i(t), x_k(t)) = \Theta(\zeta - |u^{i,j}(t) - u^{k,j}(t)|)$. Here $\Theta(...)$ is the Heaviside function.

The constructed graph is characterized by the average of all nodes' degree (<k>), where the degree, $k_i$, is the number of links formed between each node *i* and all other nodes in the system. For each node we assign $s_i = sign(\frac{\partial k_i}{\partial t})$ and then $s_i = \pm 1$. With this mapping, each node in a given time step acquires one of the states (↑ or ↓). The tuning of spins' states is achieved purely with the ultrasound waves with the form of the variable-frequency micro-wave frequency signal. In the vicinity of the transition to higher energy $<k>$ levels (*i.e*, $<k>_{max}$), deviation from the resonance frequency is controlled by the instantons frequency well-characterized by $<k>$ which is the control parameter of the system (Fig.S1-S2).

**Supplementary Information**

Supplementary information accompanies this paper.

**Acknowledgements**

This material is based upon work supported in part by the U. S. Army Research Laboratory and the U. S. Army Research Office under contract/grant number _W911NF1410276.

**Author Contributions**

All authors contributed to the analysis the results and reviewed the manuscript. H.O.G and W.A.G co-wrote the manuscript. H.O.G designed the main tests and performed the calculations. W.A.G supervised the research and helped to analyze the results.

**Competing financial interests**

The authors declare no competing financial interests.